\documentclass[sigconf, final submission, nonacm]{acmart}
\acmConference{DaSH@KDD'21}{August 15-16, 2021}{Virtual Conference}
\AtBeginDocument{%
  \providecommand\BibTeX{{%
    \normalfont B\kern-0.5em{\scshape i\kern-0.25em b}\kern-0.8em\TeX}}}

\acmSubmissionID{4}

\usepackage{multirow}

\usepackage{graphicx}
\graphicspath{ {./images/} }
\begin{document}

\title{Multi-class Text Classification using BERT-based Active Learning}

\author{Sumanth Prabhu}
\authornote{Both authors contributed equally to this research.}
\email{sumanth.prabhu@swiggy.in}
\affiliation{%
  \institution{Applied Research, Swiggy}
  \city{Bangalore}
 }
  
\author{Moosa Mohamed}
\authornotemark[1]
\email{moosa.mohamed@swiggy.in}
\affiliation{%
  \institution{Applied Research, Swiggy}
  \city{Bangalore}
}
\author{Hemant Misra}
\email{hemant.misra@swiggy.in}
\affiliation{%
  \institution{Applied Research, Swiggy}
  \city{Bangalore}
  }








\begin{abstract}
  Text classification finds interesting applications in the pickup and delivery services industry where customers require one or more items to be picked up from a location and delivered to a certain destination. Classifying these customer transactions into multiple categories helps understand the market needs for different customer segments. Each transaction is accompanied by a text description provided by the customer to describe the products being picked up and delivered which can be used to classify the transaction. BERT-based models have proven to perform well in Natural Language Understanding. However, the product descriptions provided by the customers tend to be short, incoherent and code-mixed (Hindi-English) text which demands fine-tuning of such models with manually labelled data to achieve high accuracy. Collecting this labelled data can prove to be expensive. In this paper, we explore Active Learning strategies to label transaction descriptions cost effectively while using BERT to train a transaction classification model. On TREC-6, AG's News Corpus and an internal dataset, we benchmark the performance of BERT across different Active Learning strategies in Multi-Class Text Classification.
\end{abstract}




\maketitle

\section{Introduction}
Most of the focus of the machine learning community is about creating better algorithms for learning from data. However, getting useful annotated datasets can prove to be difficult. In many scenarios, we may have access to a large unannotated corpus while it would be infeasible to annotate all instances~\citep{settles2009active}. Thus, weaker forms of supervision ~\citep{Liang05semi-supervisedlearning,theor-nips2013,snorkel-mainpaper,contextualWS2020} were explored to label data cost effectively. In this paper, we focus on Active Learning, a popular technique under Human-in-the-Loop Machine Learning~\cite{robert-munro}. Active Learning overcomes the labeling bottleneck by choosing a subset of instances from a pool of unlabeled instances to be labeled by the human annotators (a.k.a the oracle). The goal of identifying this subset is to achieve high accuracy while having labeled as few instances as possible.

BERT-based~\cite{devlin2019bert,sanh2020distilbert,roberta2019} Deep Learning models have been proven to achieve state-of-the-art results on GLUE~\citep{glue2018}, RACE~\citep{lai2017large} and SQuAD~\citep{squad2016,squad2018}. Active Learning has been successfully integrated with various Deep Learning approaches~\citep{gal2016bayesian,gissin2019discriminative}. However, the use of Active Learning with BERT based models for multi class text classification has not been studied extensively. 

In this paper, we consider a text classification use-case in industry specific to pickup and delivery services where customers make use of short text to describe the products to be picked up from a certain location and dropped at a target location. Table~\ref{table:examples} shows a few examples of the descriptions used by our customers to describe their transactions. Customers tend to use short incoherent and code-mixed (using more than one language in the same message\footnote{In our case, Hindi written in roman case is mixed with English}) textual descriptions of the products for describing them in the transaction. These descriptions, if mapped to a fixed set of possible categories, help assist critical business decisions such as demographic driven prioritization of categories, launch of new product categories and so on. Furthermore, a transaction may comprise of multiple products which adds to the complexity of the task. In this work, we focus on a multi-class classification of transactions, where a single majority category drives the transaction.

\begin{table}[bht]
\centering
\begin{tabular}{ |p{5cm}|p{2cm}| }
 \hline
 \textbf{Transaction Description} & \textbf{Category}\\
 \hline
 \textit{"Get me dahi 1.5kg"} \newline
\textbf{Translation :} Get me 1.5 kilograms of curd &  Grocery \\
\hline
\textit{"Pick up 1 yellow coloured dress"} & Clothes \\
 \hline
 \textit{"3 plate chole bhature"} &   Food\\
 \hline
  \textit{"2254/- pay krke samaan uthana hai"} \newline
  \textbf{Translation :} Pay 2254/- and pick up the package &   Package  \\
  \hline
\end{tabular}
\caption{Instances of actual transaction descriptions used by our customers along with their corresponding categories.}
\label{table:examples}
\end{table}

We explored supervised category classification of transactions which required labelled data for training. Our experiments with different approaches revealed that BERT-based models performed really well for the task. The train data used in this paper was labeled manually by subject matter experts. However, this was proving to be a very expensive exercise, and hence, necessitated exploration of cost effective strategies to collect manually labelled training data.

The key contributions of our work are as follows \begin{itemize}
\item \textbf{Active Learning with incoherent and code-mixed data:} An Active Learning framework to reduce the cost of labelling data for multi class text classification by 85\% in an industry use-case.

\item \textbf{BERT for Active Learning in multi-class text Classification} The first work, to the best of our
knowledge, to explore and compare multiple advanced strategies in Active Learning like Discriminative Active Learning using BERT for multi-class text classification on publicly available TREC-6 and AG's News Corpus benchmark datasets. 

\end{itemize}

\section{Related Work}
Active Learning has been widely studied and applied in a variety of tasks including classification~\citep{al-application-text-classification-2006,al-application-text-classification-2009}, structured output learning~\citep{al-application-structured-output-2006,al-application-structured-output-2009,al-applications-mt-2009}, clustering~\citep{pmlr-v16-bodo11a} and so on.

\subsection{Traditional Active Learning}
Popular traditional Active Learning approaches include sampling based on model uncertainty using measures such as entropy~\citep{uncertainty-sampling-1994,entropy-uncertainty-2008}, margin sampling~\citep{Scheffer01activehidden} and least model confidence~ \citep{variation-ratio-settles-2008,variation-ratio-culotta-2008,settles2009active}. Researchers also explored sampling based on diversity of instances ~\citep{mccallum-1998,variation-ratio-settles-2008,Xu_incorporatingdiversity,hierarchical-sampling-2008,submodularity-2015} to prevent selection of redundant instances for labelling. \citep{hybrid-2004,hybrid-2015,al-hybrid-2014}~explored hybrid strategies combining the advantages of uncertainty and diversity based sampling. Our work focusses on extending such strategies to Active Learning combined with Deep Learning.

\subsection{Active Learning for Deep Learning}
Deep Learning in an Active Learning setting can be challenging due to their tendency of rarely being uncertain during inference and requirement of a  relatively large amount of training data. However, Bayesian Deep Learning based approaches~\citep{gal2016bayesian,gal2017deep,kirsch2019batchbald} were leveraged to demonstrate high performance in an Active Learning setting for image classification. Similar to traditional Active Learning approaches, researchers explored diversity based sampling~\cite{sener2018active} and hybrid sampling~\cite{zhdanov2019diverse} to address the drawbacks of pure model uncertainty based approaches. Moreover, researchers also explored approaches based on Expected Model Change~\citep{egl-settles-2008,huang2016active,zhang2016active,yoo2019learning} where the selected instances are expected to result in the greatest change to the current model parameter estimates when their labels are provided. \cite{gissin2019discriminative}~ proposed ``Discriminative Active Learning (DAL)'' where Active Learning in Image Classification was posed as a binary classification task where instances to label were chosen such that the set of labeled instances and the set of unlabeled instances became indistinguishable. However, there is minimal literature in applying these strategies to Natural Language Understanding (NLU). Our work focusses on extending such Deep Active Learning strategies for Text Classification using BERT-based models.

\subsection{Deep Active Learning for Text Classification in NLP}
\citep{siddhant2018deep}~conducted an empirical study of Active Learning in NLP to observe that Bayesian Active Learning outperformed classical uncertainty sampling across all settings. However, they did not explore the performances of BERT-based models. \cite{prabhu2019sampling,Zhang2019AnED}~explored Active Learning strategies extensible to BERT-based models. \cite{prabhu2019sampling}~conducted an empirical study to demonstrate that active set selection using the posterior entropy of deep models like FastText.zip (FTZ) is robust to sampling biases and to various algorithmic choices such as BERT. \cite{Zhang2019AnED}~applied an ensemble
of Active Learning strategies to BERT for the task of intent classification. However, neither of these works account for comparison against advanced strategies like Discriminative Active Learning. \cite{ein-dor-etal-2020-active}~explored Active Learning with multiple strategies using BERT based models for binary text classification tasks. Our work is very closely related to this work with the difference that we focus on experiments with multi-class text classification.




\section{Methodology}
\begin{figure*}[h]
\centering
\includegraphics[width=15cm, trim={0cm 0cm 0.5cm 2cm}, ]{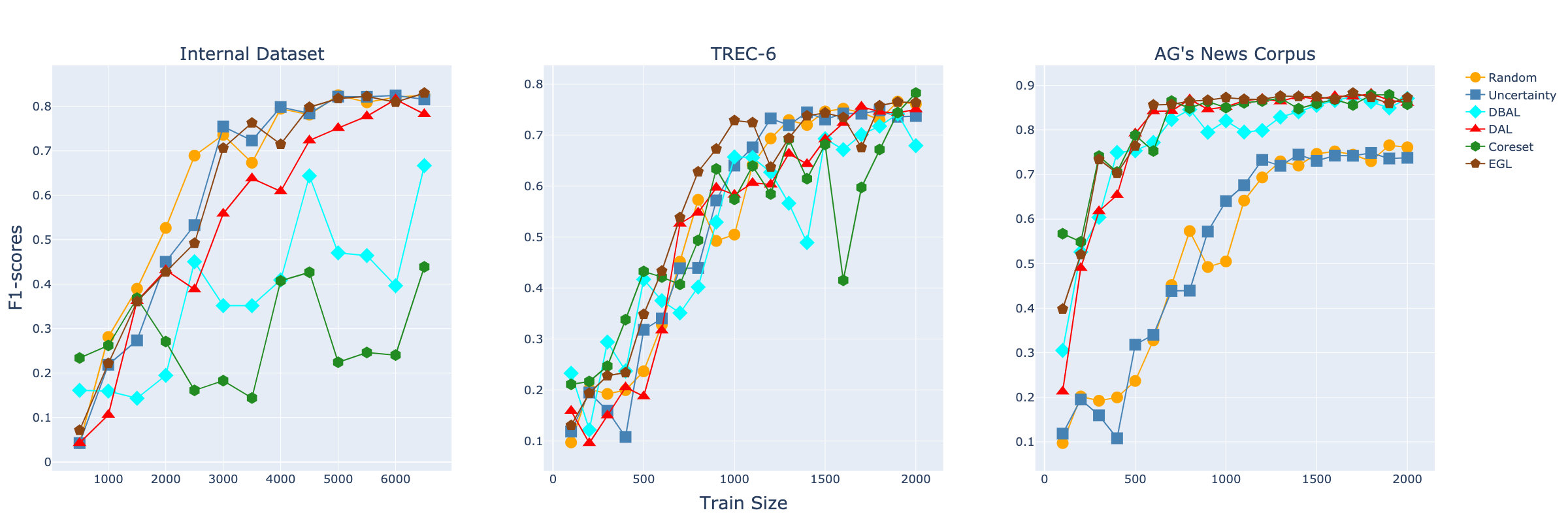}
\caption{Comparison of F1 scores of various Active Learning strategies on the Internal Dataset(BatchSize=500, Iterations=13), TREC-6 Dataset(BatchSize=100, Iterations=20) and AG's News Corpus(BatchSize=100, Iterations=20)}
\label{fig:F1-scores-large-batchsize}
\end{figure*}
We considered multiple strategies in Active Learning to understand the impact of the performance using BERT based models. We attempted to reproduce settings similar to the study conducted by Ein-Dor et al.~\citep{ein-dor-etal-2020-active} for binary text classification.
\begin{itemize}
  \item \textbf{Random} We sampled instances at random in each iteration while leveraging them to retrain the model.
  \item \textbf{Uncertainty (Entropy)} Instances selected to retrain the model had the highest entropy in model predictions.
  \item \textbf{Expected Gradient Length (EGL)}~\citep{huang2016active} This strategy picked instances that are expected to have the largest gradient norm over all possible labelings of the instances.
  \item \textbf{Deep Bayesian Active Learning (DBAL)}~\citep{gal2017deep} Instances were selected to improve the uncertainty measures of BERT with Monte Carlo dropout using the max-entropy acquisition function.
  \item \textbf{Core-set}~\citep{sener2018active} selected instances that best cover the dataset in the learned representation space using farthest-first traversal algorithm
  \item \textbf{Discriminative Active Learning (DAL)}~\citep{gissin2019discriminative} This strategy aimed to select instances that make the labeled set of instances indistinguishable from the unlabeled pool.
\end{itemize}

\section{Experiments}

\subsection{Data}
Our \textbf{Internal Dataset} comprised of 55,038 customer transactions each of which had an associated transaction description  provided  by  the  customer.  The instances  were  manually  annotated  by  a  team  of SMEs and mapped to one of 10 pre-defined categories.  This data set was split into 44,030 training samples and 11,008 validation samples. The list of categories considered are as follows: \{`Food’, `Grocery', `Package', `Medicines', `Household Items', `Cigarettes', `Clothes', `Electronics', `Keys', `Documents/Books' \} 
\hfil \break
To verify the reproducibility of our observations, we considered the \textbf{TREC dataset}~\citep{data-trec6-2002} which comprised of 5,452 training samples and 500 validation samples with 6 classes and also considered the \textbf{AG’s News Corpus}~\citep{data-ag-2016} which comprised of 120,000 training samples and 7,600 validation samples with 4 classes.


\subsection{Experiment Setting}
 
We leveraged DistilBERT~\citep{sanh2020distilbert} for the purpose of the experiments in this paper. We ran each strategy for two random seed settings on the TREC-6 Dataset and AG's News Corpus and one random seed setting for our Internal Dataset.  The results presented in the paper consist of 558 fine-tuning experiments (78 for our Internal dataset and 240 each for the TREC-6 dataset and AG's News Corpus). The experiments were run on AWS Sagemaker's p3 2x large Spot Instances. The implementation was based on the code\footnote{https://github.com/dsgissin/DiscriminativeActiveLearning} made available by~\cite{gissin2019discriminative}. For the our Internal Dataset, we set the batch size to 500 and number of iterations to 13. For TREC-6 and AG's News Corpus, we set the batch size and number of iterations to 100 and 20 respectively. In each iteration, a batch of samples were identified and the model was retrained for 1 epoch with a learning rate of $3 \times 10^{-5}$. We retrained the models from scratch in each iteration to prevent overfitting~\citep{siddhant2018deep,hu2019active}. 






\section{Results}


We report results for the Active Learning Strategies using F1-score as the classification metric. Figure~\ref{fig:F1-scores-large-batchsize} visualizes the performance of different approaches. We observe that the best F1-score that Active Learning using BERT achieves on the Internal Dataset is 0.83. A fully supervised setting with DistilBERT using the entire training data also achieved an F1-score of 0.83. Thus, we reduce the labelling cost by 85\% while achieving similar F1-scores.  However, we observe that no single strategy significantly outperforms the other. \citep{lowell2019practical}~who studied Active Learning for text classification and sequence tagging in non-BERT models, and demonstrated the brittleness
and inconsistency of Active Learning results. \citep{ein-dor-etal-2020-active}~also observed inconsistency of results using BERT in binary classification. Moreover, we observe that the performance improvements over Random Sampling strategy is also inconsistent. Concretely, on the Internal Dataset and TREC-6, we observe that DBAL and Core-set are performing poorly when compared to Random Sampling. However, on the AG's News Corpus, we observe a contradicting behaviour where both DBAL and Core-set outperform the Random Sampling strategy. Uncertainty sampling strategy performs very similar to the Random Sampling strategy. EGL performs consistently well across all three datasets. The DAL strategy, proven to have worked well in image classification, is consistent across the datasets. However, it does not outperform the EGL strategy.




\section{Analysis}

In this section, we analyze the different AL strategies considered in the paper to understand their relative advantages and disadvantages. Concretely, we consider the following metrics - 
\begin{itemize}
    \item \textbf{Diversity} The instances chosen in each iteration need to be as different from each other as possible. We rely on the Diversity measure proposed by ~\citep{ein-dor-etal-2020-active} based on the euclidean distance between the [CLS] represetations of the samples.
    
\begin{table}[b]
\centering
\begin{tabular}{|p{1.4cm}|p{1.5cm}|p{1.3cm}|p{2.8cm}|}
\hline
\textbf{Dataset} & \textbf{Approach} & \textbf{Diversity} & \textbf{Representativeness} \\ \hline
 & Random & 0.28 & 0.37 \\ \cline{2-4}
 & Uncertainty & 0.25 & 0.34 \\ \cline{2-4}
\multirow{5}{\linewidth}{TREC-6} & Core-set & 0.25 & 0.31 \\ \cline{2-4}
 & DAL & 0.26 & 0.35 \\ \cline{2-4}
 & DBAL & 0.25 & 0.33 \\ \cline{2-4}
 & EGL & 0.27 & 0.35 \\ \hline
 & Random & 0.26 & 0.43 \\ \cline{2-4}
 & Uncertainty & 0.21 & 0.37\\ \cline{2-4}
\multirow{5}{\linewidth}{AG's News Corpus} & Core-set & 0.24 & 0.39 \\ \cline{2-4}
 & DAL & 0.25 & 0.4 \\ \cline{2-4}
 & DBAL & 0.23 & 0.39 \\ \cline{2-4}
 & EGL & 0.24 & 0.39 \\ \hline

\end{tabular}
\caption{Diversity and Representativeness for the TREC-6 Dataset and AG's News Corpus}
\label{table:diversty_representativeness}
\end{table}
    \item \textbf{Representativeness} The AL strategies that select highly diverse instances can still have a tendency to select outlier instances that are not representative of the overall data distribution. Following ~\citep{ein-dor-etal-2020-active}, we capture the representativeness of the selected instances using the KNN-density measure. We quantify the density of an instance as one over the average distance between the selected instances and their K nearest neighbors within the unlabelled pool, based on the [CLS] representations.
    


    \item \textbf{Class Bias}  Inspired by~\cite{prabhu2019sampling}, we consider quantifying the level of disproportion in the class labels for each of the the AL strategies. Concretely, we measure potential Class Bias using Label Entropy~\cite{prabhu2019sampling} based Kullback-Leibler (KL) divergence between the ground-truth label distribution and the distribution obtained from the chosen instances. 
    
    \begin{table}[b]
\centering
\begin{tabular}{|p{1.1cm}|p{1.6cm}|p{1.7cm}|p{1.7cm}|}
\hline
\textbf{Dataset} & \textbf{Approach} & \textbf{$\cap$ Q} & \textbf{$\cap$ S} \\ \hline
 & Random & $1.79 \pm0.01$ & $1.79$ \\ \cline{2-4}
 & Uncertainty & $1.26 \pm0.23$ & $1.76$ \\ \cline{2-4}
\multirow{5}{\linewidth}{TREC-6} & Core-set & $1.64 \pm 0.15$ & $1.77$\\ \cline{2-4}
 & DAL & $1.75 \pm 0.02$ & $1.79$ \\ \cline{2-4}
 & DBAL & $1.38 \pm 0.09$ & $1.75$ \\ \cline{2-4}
 & EGL & $1.75 \pm 0.02$ & $1.79$ \\ \hline
 & Random & $1.37 \pm 0.01$ & $1.39$ \\ \cline{2-4}
 & Uncertainty & $0.87 \pm 0.43$ & $1.36$ \\ \cline{2-4}
\multirow{5}{\linewidth}{AG's News Corpus} & Core-set & $0.98 \pm 0.42$ & $1.22$ \\ \cline{2-4}
 & DAL & $1.33 \pm 0.03$ & $1.38$ \\ \cline{2-4}
 & DBAL & $0.78 \pm 0.31$ & $1.38$ \\ \cline{2-4}
 & EGL & $1.18 \pm 0.05$ & $1.35$ \\ \hline

\end{tabular}
\caption{Label Entropies for the TREC-6 Dataset and AG's News Corpus. $\cap$ Q denotes averaging across queries of a single run, $\cap$ S denotes the label entropy of the final collected samples}
\label{table:class-bias}
\end{table}

\begin{table}[h]
\centering
\begin{tabular}{|p{2.0cm}|p{3.0cm}|}
\hline
\textbf{Approach} & \textbf{Runtime (seconds)} \\ \hline
Random & <1 \\ \hline
Uncertainty & 84 \\ \hline
Core-set & 89 \\ \hline
DAL & 378 \\ \hline
DBAL & 120 \\ \hline
EGL & 484 \\  \hline
\end{tabular}
\caption{Runtimes (in seconds) for the TREC-6 Dataset per iteration for different AL strategies, with 5,000 samples}
\label{table:run-times}
\end{table}    
    
\begin{figure}[h]
\centering
\includegraphics[width=7cm, trim={0cm 0cm 0cm 0cm}, ]{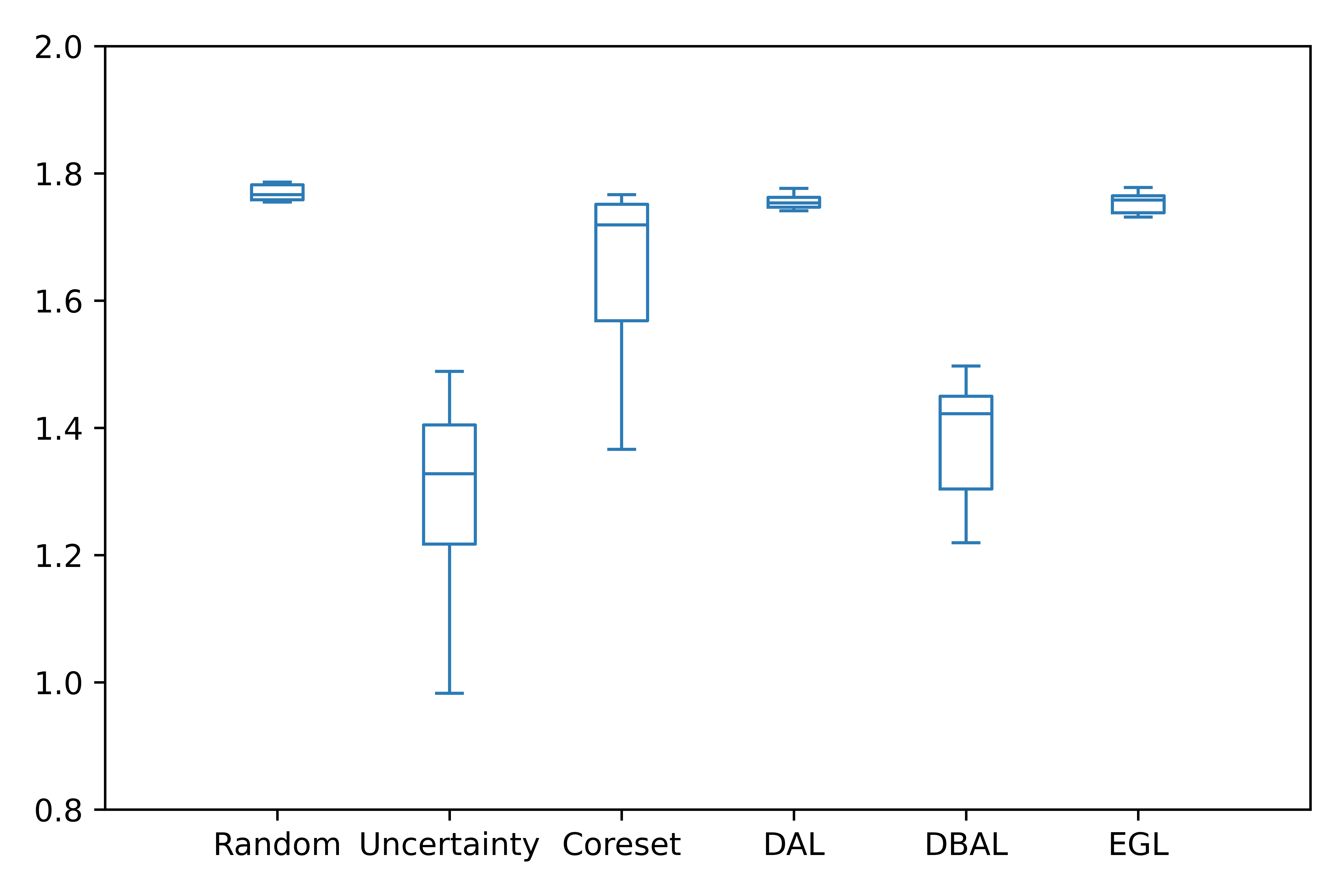}
\includegraphics[width=6.9cm, trim={0cm 0cm 0cm 0cm}, ]{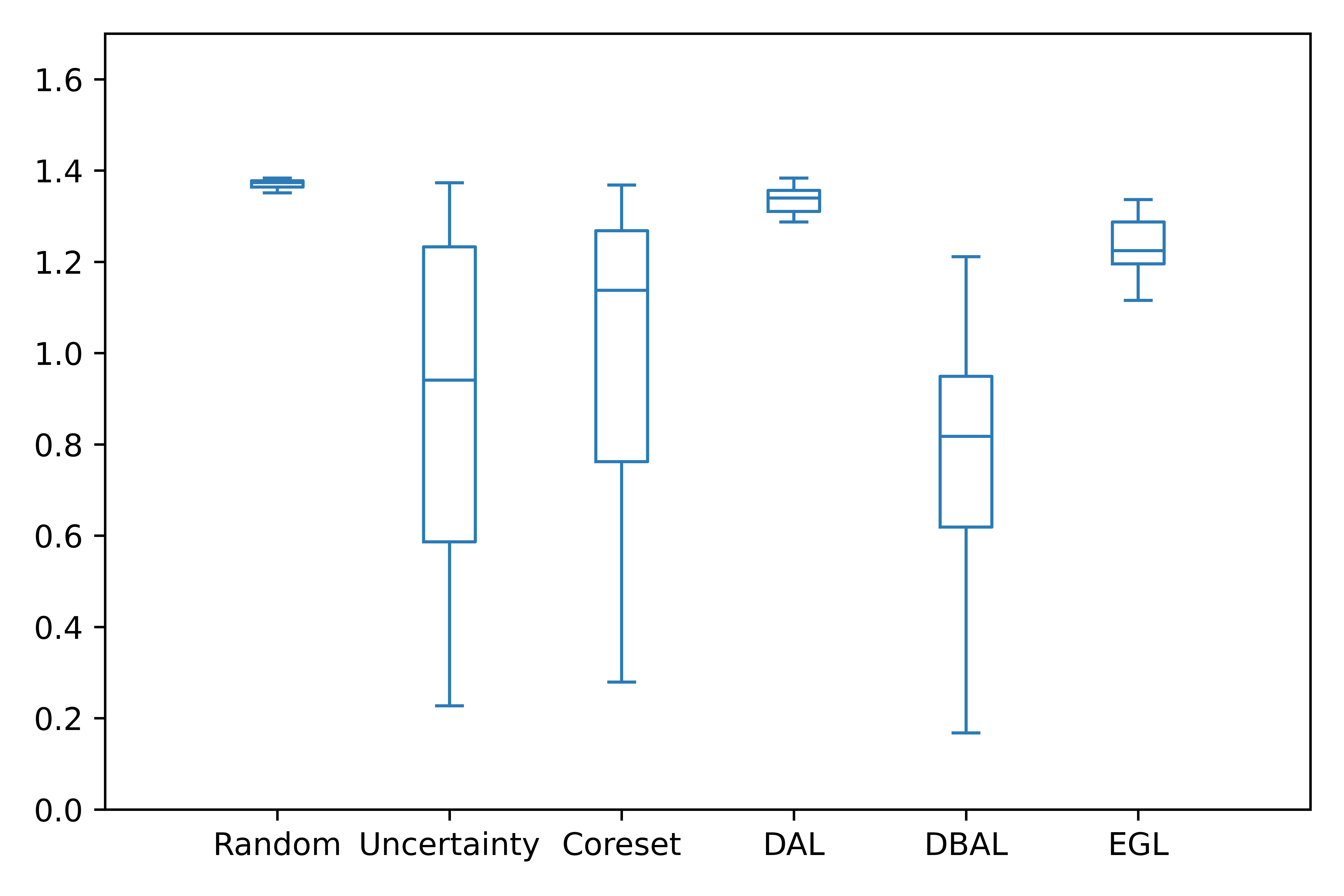}
\caption{Class Bias scores per run for TREC-6 (top) and AG's News (bottom) across different AL strategies}
\label{fig:class-bias}

\end{figure}


    \item \textbf{Runtime} Another important factor for comparison of different AL strategies is the time taken to execute each iteration. We compare the efficiency of the AL strategies based on runtimes.

\end{itemize}

Table~ \ref{table:diversty_representativeness} shows the  diversity and representativeness of different strategies for the datasets. We observed that Random achieved the highest scores while DAL and EGL achieved relatively higher scores compared to the remaining strategies. Core-set, which was designed to achieve high diversity, also achieves high scores in Diversity for both datasets. However, we observe that it displays inconsistent behaviour in selecting representative instances which could potentially be attributed to its tendency to select outliers. Uncertainty achieves the lowest score in diversity and is also inconsistent in selecting representative instances. Interpreting the scores would require a deeper analysis of these observations which we leave to future work.

Table~\ref{table:class-bias} shows the class bias with different approaches. Figure~\ref{fig:class-bias} shows the corresponding boxplots for the AL strategies. Wilcoxon signed-rank test~\cite{Rey2011} showed that Random, DAL and EGL have significantly higher label entropies, thus, lower class bias compared to the remaining strategies for the datasets considered in the paper. However, DAL and EGL can be relatively time consuming. Table~\ref{table:run-times} shows the runtimes for each AL approach. We observed that Random is the fastest strategy while EGL is the slowest owing to its dependency on the gradient calculation.

\section{Conclusion}
In this paper, we explored multiple Active Learning strategies using BERT. Our goal was to understand if BERT-based models can prove effective in an Active Learning setting for multi-class text classification. We observed that EGL performed reasonably well across datasets in multi-class text classification. Moreover, Random, EGL and DAL captured diverse and representative samples with relatively lower class bias. However, unlike Random, EGL and DAL had longer execution times. In future work, we plan to perform a deeper analysis of the observations and also explore ensembling approaches by combining the advantages of each strategy to explore potential performance improvements. 

\bibliographystyle{ACM-Reference-Format}
\bibliography{sample-base}


\begin{thebibliography}{50}


\ifx \showCODEN    \undefined \def \showCODEN     #1{\unskip}     \fi
\ifx \showDOI      \undefined \def \showDOI       #1{#1}\fi
\ifx \showISBNx    \undefined \def \showISBNx     #1{\unskip}     \fi
\ifx \showISBNxiii \undefined \def \showISBNxiii  #1{\unskip}     \fi
\ifx \showISSN     \undefined \def \showISSN      #1{\unskip}     \fi
\ifx \showLCCN     \undefined \def \showLCCN      #1{\unskip}     \fi
\ifx \shownote     \undefined \def \shownote      #1{#1}          \fi
\ifx \showarticletitle \undefined \def \showarticletitle #1{#1}   \fi
\ifx \showURL      \undefined \def \showURL       {\relax}        \fi
\providecommand\bibfield[2]{#2}
\providecommand\bibinfo[2]{#2}
\providecommand\natexlab[1]{#1}
\providecommand\showeprint[2][]{arXiv:#2}

\bibitem[\protect\citeauthoryear{Baram, El-Yaniv, and Luz}{Baram
  et~al\mbox{.}}{2004}]%
        {hybrid-2004}
\bibfield{author}{\bibinfo{person}{Yoram Baram}, \bibinfo{person}{Ran
  El-Yaniv}, {and} \bibinfo{person}{Kobi Luz}.}
  \bibinfo{year}{2004}\natexlab{}.
\newblock \showarticletitle{Online Choice of Active Learning Algorithms}.
\newblock \bibinfo{journal}{\emph{J. Mach. Learn. Res.}}  \bibinfo{volume}{5}
  (\bibinfo{date}{Dec.} \bibinfo{year}{2004}), \bibinfo{pages}{255–291}.
\newblock
\showISSN{1532-4435}


\bibitem[\protect\citeauthoryear{Bodó, Minier, and Csató}{Bodó
  et~al\mbox{.}}{2011}]%
        {pmlr-v16-bodo11a}
\bibfield{author}{\bibinfo{person}{Zalán Bodó}, \bibinfo{person}{Zsolt
  Minier}, {and} \bibinfo{person}{Lehel Csató}.}
  \bibinfo{year}{2011}\natexlab{}.
\newblock \showarticletitle{Active Learning with Clustering}. In
  \bibinfo{booktitle}{\emph{Active Learning and Experimental Design workshop In
  conjunction with AISTATS 2010}} \emph{(\bibinfo{series}{Proceedings of
  Machine Learning Research}, Vol.~\bibinfo{volume}{16})},
  \bibfield{editor}{\bibinfo{person}{Isabelle Guyon}, \bibinfo{person}{Gavin
  Cawley}, \bibinfo{person}{Gideon Dror}, \bibinfo{person}{Vincent Lemaire},
  {and} \bibinfo{person}{Alexander Statnikov}} (Eds.). \bibinfo{publisher}{JMLR
  Workshop and Conference Proceedings}, \bibinfo{address}{Sardinia, Italy},
  \bibinfo{pages}{127--139}.
\newblock
\urldef\tempurl%
\url{http://proceedings.mlr.press/v16/bodo11a.html}
\showURL{%
\tempurl}


\bibitem[\protect\citeauthoryear{Culotta and McCallum}{Culotta and
  McCallum}{2005}]%
        {variation-ratio-culotta-2008}
\bibfield{author}{\bibinfo{person}{Aron Culotta} {and} \bibinfo{person}{Andrew
  McCallum}.} \bibinfo{year}{2005}\natexlab{}.
\newblock \showarticletitle{Reducing Labeling Effort for Structured Prediction
  Tasks}. In \bibinfo{booktitle}{\emph{Proceedings of the 20th National
  Conference on Artificial Intelligence - Volume 2}} (Pittsburgh, Pennsylvania)
  \emph{(\bibinfo{series}{AAAI'05})}. \bibinfo{publisher}{AAAI Press},
  \bibinfo{pages}{746–751}.
\newblock
\showISBNx{157735236x}


\bibitem[\protect\citeauthoryear{Dasgupta and Hsu}{Dasgupta and Hsu}{2008}]%
        {hierarchical-sampling-2008}
\bibfield{author}{\bibinfo{person}{Sanjoy Dasgupta} {and}
  \bibinfo{person}{Daniel Hsu}.} \bibinfo{year}{2008}\natexlab{}.
\newblock \showarticletitle{Hierarchical Sampling for Active Learning}
  \emph{(\bibinfo{series}{ICML'08})}. \bibinfo{publisher}{Association for
  Computing Machinery}, \bibinfo{address}{New York, NY, USA},
  \bibinfo{pages}{208–215}.
\newblock
\showISBNx{9781605582054}
\urldef\tempurl%
\url{https://doi.org/10.1145/1390156.1390183}
\showDOI{\tempurl}


\bibitem[\protect\citeauthoryear{Devlin, Chang, Lee, and Toutanova}{Devlin
  et~al\mbox{.}}{2019}]%
        {devlin2019bert}
\bibfield{author}{\bibinfo{person}{Jacob Devlin}, \bibinfo{person}{Ming-Wei
  Chang}, \bibinfo{person}{Kenton Lee}, {and} \bibinfo{person}{Kristina
  Toutanova}.} \bibinfo{year}{2019}\natexlab{}.
\newblock \showarticletitle{{BERT}: Pre-training of Deep Bidirectional
  Transformers for Language Understanding}. In
  \bibinfo{booktitle}{\emph{Proceedings of the 2019 Conference of the North
  {A}merican Chapter of the Association for Computational Linguistics: Human
  Language Technologies, Volume 1 (Long and Short Papers)}}.
  \bibinfo{publisher}{Association for Computational Linguistics},
  \bibinfo{address}{Minneapolis, Minnesota}, \bibinfo{pages}{4171--4186}.
\newblock
\urldef\tempurl%
\url{https://doi.org/10.18653/v1/N19-1423}
\showDOI{\tempurl}


\bibitem[\protect\citeauthoryear{Ein-Dor, Halfon, Gera, Shnarch, Dankin,
  Choshen, Danilevsky, Aharonov, Katz, and Slonim}{Ein-Dor
  et~al\mbox{.}}{2020}]%
        {ein-dor-etal-2020-active}
\bibfield{author}{\bibinfo{person}{Liat Ein-Dor}, \bibinfo{person}{Alon
  Halfon}, \bibinfo{person}{Ariel Gera}, \bibinfo{person}{Eyal Shnarch},
  \bibinfo{person}{Lena Dankin}, \bibinfo{person}{Leshem Choshen},
  \bibinfo{person}{Marina Danilevsky}, \bibinfo{person}{Ranit Aharonov},
  \bibinfo{person}{Yoav Katz}, {and} \bibinfo{person}{Noam Slonim}.}
  \bibinfo{year}{2020}\natexlab{}.
\newblock \showarticletitle{{A}ctive {L}earning for {BERT}: {A}n {E}mpirical
  {S}tudy}. In \bibinfo{booktitle}{\emph{Proceedings of the 2020 Conference on
  Empirical Methods in Natural Language Processing (EMNLP)}}.
  \bibinfo{publisher}{Association for Computational Linguistics},
  \bibinfo{address}{Online}, \bibinfo{pages}{7949--7962}.
\newblock
\urldef\tempurl%
\url{https://doi.org/10.18653/v1/2020.emnlp-main.638}
\showDOI{\tempurl}


\bibitem[\protect\citeauthoryear{Gal and Ghahramani}{Gal and
  Ghahramani}{2016}]%
        {gal2016bayesian}
\bibfield{author}{\bibinfo{person}{Yarin Gal} {and} \bibinfo{person}{Zoubin
  Ghahramani}.} \bibinfo{year}{2016}\natexlab{}.
\newblock \bibinfo{title}{Bayesian Convolutional Neural Networks with Bernoulli
  Approximate Variational Inference}.
\newblock
\newblock
\showeprint[arxiv]{1506.02158}~[stat.ML]


\bibitem[\protect\citeauthoryear{Gal, Islam, and Ghahramani}{Gal
  et~al\mbox{.}}{2017}]%
        {gal2017deep}
\bibfield{author}{\bibinfo{person}{Yarin Gal}, \bibinfo{person}{Riashat Islam},
  {and} \bibinfo{person}{Zoubin Ghahramani}.} \bibinfo{year}{2017}\natexlab{}.
\newblock \showarticletitle{Deep {B}ayesian Active Learning with Image Data}.
  In \bibinfo{booktitle}{\emph{Proceedings of the 34th International Conference
  on Machine Learning}} \emph{(\bibinfo{series}{Proceedings of Machine Learning
  Research}, Vol.~\bibinfo{volume}{70})},
  \bibfield{editor}{\bibinfo{person}{Doina Precup} {and}
  \bibinfo{person}{Yee~Whye Teh}} (Eds.). \bibinfo{publisher}{PMLR},
  \bibinfo{pages}{1183--1192}.
\newblock
\urldef\tempurl%
\url{http://proceedings.mlr.press/v70/gal17a.html}
\showURL{%
\tempurl}


\bibitem[\protect\citeauthoryear{Gissin and Shalev-Shwartz}{Gissin and
  Shalev-Shwartz}{2019}]%
        {gissin2019discriminative}
\bibfield{author}{\bibinfo{person}{Daniel Gissin} {and} \bibinfo{person}{Shai
  Shalev-Shwartz}.} \bibinfo{year}{2019}\natexlab{}.
\newblock \bibinfo{title}{Discriminative Active Learning}.
\newblock
\newblock
\showeprint[arxiv]{1907.06347}~[cs.LG]


\bibitem[\protect\citeauthoryear{Haffari and Sarkar}{Haffari and
  Sarkar}{2009}]%
        {al-applications-mt-2009}
\bibfield{author}{\bibinfo{person}{Gholamreza Haffari} {and}
  \bibinfo{person}{Anoop Sarkar}.} \bibinfo{year}{2009}\natexlab{}.
\newblock \showarticletitle{Active Learning for Multilingual Statistical
  Machine Translation}. In \bibinfo{booktitle}{\emph{Proceedings of the Joint
  Conference of the 47th Annual Meeting of the {ACL} and the 4th International
  Joint Conference on Natural Language Processing of the {AFNLP}}}.
  \bibinfo{publisher}{Association for Computational Linguistics},
  \bibinfo{address}{Suntec, Singapore}, \bibinfo{pages}{181--189}.
\newblock
\urldef\tempurl%
\url{https://www.aclweb.org/anthology/P09-1021}
\showURL{%
\tempurl}


\bibitem[\protect\citeauthoryear{Hsu and Lin}{Hsu and Lin}{2015}]%
        {hybrid-2015}
\bibfield{author}{\bibinfo{person}{Wei-Ning Hsu} {and}
  \bibinfo{person}{Hsuan-Tien Lin}.} \bibinfo{year}{2015}\natexlab{}.
\newblock \showarticletitle{Active Learning by Learning}. In
  \bibinfo{booktitle}{\emph{Proceedings of the Twenty-Ninth AAAI Conference on
  Artificial Intelligence}} (Austin, Texas) \emph{(\bibinfo{series}{AAAI'15})}.
  \bibinfo{publisher}{AAAI Press}, \bibinfo{pages}{2659–2665}.
\newblock
\showISBNx{0262511290}


\bibitem[\protect\citeauthoryear{Hu, Lipton, Anandkumar, and Ramanan}{Hu
  et~al\mbox{.}}{2019}]%
        {hu2019active}
\bibfield{author}{\bibinfo{person}{Peiyun Hu}, \bibinfo{person}{Zachary~C.
  Lipton}, \bibinfo{person}{Anima Anandkumar}, {and} \bibinfo{person}{Deva
  Ramanan}.} \bibinfo{year}{2019}\natexlab{}.
\newblock \bibinfo{title}{Active Learning with Partial Feedback}.
\newblock
\newblock
\showeprint[arxiv]{1802.07427}~[cs.LG]


\bibitem[\protect\citeauthoryear{Huang, Child, Rao, Liu, Satheesh, and
  Coates}{Huang et~al\mbox{.}}{2016}]%
        {huang2016active}
\bibfield{author}{\bibinfo{person}{Jiaji Huang}, \bibinfo{person}{Rewon Child},
  \bibinfo{person}{Vinay Rao}, \bibinfo{person}{Hairong Liu},
  \bibinfo{person}{Sanjeev Satheesh}, {and} \bibinfo{person}{Adam Coates}.}
  \bibinfo{year}{2016}\natexlab{}.
\newblock \bibinfo{title}{Active Learning for Speech Recognition: the Power of
  Gradients}.
\newblock
\newblock
\showeprint[arxiv]{1612.03226}~[cs.CL]


\bibitem[\protect\citeauthoryear{Kirsch, van Amersfoort, and Gal}{Kirsch
  et~al\mbox{.}}{2019}]%
        {kirsch2019batchbald}
\bibfield{author}{\bibinfo{person}{Andreas Kirsch}, \bibinfo{person}{Joost~R.
  van Amersfoort}, {and} \bibinfo{person}{Y. Gal}.}
  \bibinfo{year}{2019}\natexlab{}.
\newblock \showarticletitle{BatchBALD: Efficient and Diverse Batch Acquisition
  for Deep Bayesian Active Learning}. In \bibinfo{booktitle}{\emph{NeurIPS}}.
\newblock


\bibitem[\protect\citeauthoryear{Lai, Xie, Liu, Yang, and Hovy}{Lai
  et~al\mbox{.}}{2017}]%
        {lai2017large}
\bibfield{author}{\bibinfo{person}{Guokun Lai}, \bibinfo{person}{Qizhe Xie},
  \bibinfo{person}{Hanxiao Liu}, \bibinfo{person}{Yiming Yang}, {and}
  \bibinfo{person}{Eduard Hovy}.} \bibinfo{year}{2017}\natexlab{}.
\newblock \showarticletitle{RACE: Large-scale ReAding Comprehension Dataset
  From Examinations}.
\newblock \bibinfo{journal}{\emph{arXiv preprint arXiv:1704.04683}}
  (\bibinfo{year}{2017}).
\newblock


\bibitem[\protect\citeauthoryear{Lewis and Gale}{Lewis and Gale}{1994}]%
        {uncertainty-sampling-1994}
\bibfield{author}{\bibinfo{person}{David~D. Lewis} {and}
  \bibinfo{person}{William~A. Gale}.} \bibinfo{year}{1994}\natexlab{}.
\newblock \showarticletitle{A Sequential Algorithm for Training Text
  Classifiers}.
\newblock \bibinfo{journal}{\emph{CoRR}}  \bibinfo{volume}{abs/cmp-lg/9407020}
  (\bibinfo{year}{1994}).
\newblock
\showeprint[arxiv]{cmp-lg/9407020}
\urldef\tempurl%
\url{http://arxiv.org/abs/cmp-lg/9407020}
\showURL{%
\tempurl}


\bibitem[\protect\citeauthoryear{Li and Roth}{Li and Roth}{2002}]%
        {data-trec6-2002}
\bibfield{author}{\bibinfo{person}{Xin Li} {and} \bibinfo{person}{Dan Roth}.}
  \bibinfo{year}{2002}\natexlab{}.
\newblock \showarticletitle{Learning Question Classifiers}. In
  \bibinfo{booktitle}{\emph{Proceedings of the 19th International Conference on
  Computational Linguistics - Volume 1}} (Taipei, Taiwan)
  \emph{(\bibinfo{series}{COLING '02})}. \bibinfo{publisher}{Association for
  Computational Linguistics}, \bibinfo{address}{USA}, \bibinfo{pages}{1–7}.
\newblock
\urldef\tempurl%
\url{https://doi.org/10.3115/1072228.1072378}
\showDOI{\tempurl}


\bibitem[\protect\citeauthoryear{Liang}{Liang}{2005}]%
        {Liang05semi-supervisedlearning}
\bibfield{author}{\bibinfo{person}{Percy Liang}.}
  \bibinfo{year}{2005}\natexlab{}.
\newblock \showarticletitle{Semi-supervised learning for natural language}. In
  \bibinfo{booktitle}{\emph{MASTER’S THESIS, MIT}}.
\newblock


\bibitem[\protect\citeauthoryear{Lowell, Lipton, and Wallace}{Lowell
  et~al\mbox{.}}{2019}]%
        {lowell2019practical}
\bibfield{author}{\bibinfo{person}{David Lowell},
  \bibinfo{person}{Zachary~Chase Lipton}, {and} \bibinfo{person}{Byron~C.
  Wallace}.} \bibinfo{year}{2019}\natexlab{}.
\newblock \showarticletitle{Practical Obstacles to Deploying Active Learning}.
  In \bibinfo{booktitle}{\emph{EMNLP/IJCNLP}}.
\newblock


\bibitem[\protect\citeauthoryear{McCallum and Nigam}{McCallum and
  Nigam}{1998}]%
        {mccallum-1998}
\bibfield{author}{\bibinfo{person}{Andrew McCallum} {and}
  \bibinfo{person}{Kamal Nigam}.} \bibinfo{year}{1998}\natexlab{}.
\newblock \showarticletitle{Employing EM and Pool-Based Active Learning for
  Text Classification}. In \bibinfo{booktitle}{\emph{Proceedings of the
  Fifteenth International Conference on Machine Learning}}
  \emph{(\bibinfo{series}{ICML '98})}. \bibinfo{publisher}{Morgan Kaufmann
  Publishers Inc.}, \bibinfo{address}{San Francisco, CA, USA},
  \bibinfo{pages}{350–358}.
\newblock
\showISBNx{1558605568}


\bibitem[\protect\citeauthoryear{Mekala and Shang}{Mekala and Shang}{2020}]%
        {contextualWS2020}
\bibfield{author}{\bibinfo{person}{Dheeraj Mekala} {and}
  \bibinfo{person}{Jingbo Shang}.} \bibinfo{year}{2020}\natexlab{}.
\newblock \showarticletitle{Contextualized Weak Supervision for Text
  Classification}. In \bibinfo{booktitle}{\emph{Association for Computational
  Linguistics}} \emph{(\bibinfo{series}{ACL'20})}.
\newblock


\bibitem[\protect\citeauthoryear{Monarch}{Monarch}{2019}]%
        {robert-munro}
\bibfield{author}{\bibinfo{person}{Robert~(Munro) Monarch}.}
  \bibinfo{year}{2019}\natexlab{}.
\newblock \bibinfo{booktitle}{\emph{Human-in-the-Loop Machine Learning}}.
\newblock \bibinfo{publisher}{Manning Publications}.
\newblock


\bibitem[\protect\citeauthoryear{Natarajan, Dhillon, Ravikumar, and
  Tewari}{Natarajan et~al\mbox{.}}{2013}]%
        {theor-nips2013}
\bibfield{author}{\bibinfo{person}{Nagarajan Natarajan},
  \bibinfo{person}{Inderjit~S. Dhillon}, \bibinfo{person}{Pradeep Ravikumar},
  {and} \bibinfo{person}{Ambuj Tewari}.} \bibinfo{year}{2013}\natexlab{}.
\newblock \showarticletitle{Learning with Noisy Labels}. In
  \bibinfo{booktitle}{\emph{Neural Information Processing Systems}}
  \emph{(\bibinfo{series}{NIPS'13})}.
\newblock


\bibitem[\protect\citeauthoryear{Novak, Mladeni{\v{c}}, and Grobelnik}{Novak
  et~al\mbox{.}}{2006}]%
        {al-application-text-classification-2006}
\bibfield{author}{\bibinfo{person}{Bla{\v{z}} Novak}, \bibinfo{person}{Dunja
  Mladeni{\v{c}}}, {and} \bibinfo{person}{Marko Grobelnik}.}
  \bibinfo{year}{2006}\natexlab{}.
\newblock \showarticletitle{Text Classification with Active Learning}. In
  \bibinfo{booktitle}{\emph{From Data and Information Analysis to Knowledge
  Engineering}}, \bibfield{editor}{\bibinfo{person}{Myra Spiliopoulou},
  \bibinfo{person}{Rudolf Kruse}, \bibinfo{person}{Christian Borgelt},
  \bibinfo{person}{Andreas N{\"u}rnberger}, {and} \bibinfo{person}{Wolfgang
  Gaul}} (Eds.). \bibinfo{publisher}{Springer Berlin Heidelberg},
  \bibinfo{address}{Berlin, Heidelberg}, \bibinfo{pages}{398--405}.
\newblock
\showISBNx{978-3-540-31314-4}


\bibitem[\protect\citeauthoryear{Prabhu, Dognin, and Singh}{Prabhu
  et~al\mbox{.}}{2019}]%
        {prabhu2019sampling}
\bibfield{author}{\bibinfo{person}{Ameya Prabhu}, \bibinfo{person}{Charles
  Dognin}, {and} \bibinfo{person}{Maneesh Singh}.}
  \bibinfo{year}{2019}\natexlab{}.
\newblock \showarticletitle{Sampling Bias in Deep Active Classification: An
  Empirical Study}. In \bibinfo{booktitle}{\emph{Proceedings of the 2019
  Conference on Empirical Methods in Natural Language Processing and the 9th
  International Joint Conference on Natural Language Processing
  (EMNLP-IJCNLP)}}. \bibinfo{publisher}{Association for Computational
  Linguistics}, \bibinfo{address}{Hong Kong, China},
  \bibinfo{pages}{4058--4068}.
\newblock
\urldef\tempurl%
\url{https://doi.org/10.18653/v1/D19-1417}
\showDOI{\tempurl}


\bibitem[\protect\citeauthoryear{Rajpurkar, Jia, and Liang}{Rajpurkar
  et~al\mbox{.}}{2018}]%
        {squad2018}
\bibfield{author}{\bibinfo{person}{Pranav Rajpurkar}, \bibinfo{person}{Robin
  Jia}, {and} \bibinfo{person}{Percy Liang}.} \bibinfo{year}{2018}\natexlab{}.
\newblock \showarticletitle{Know what you don’t know: Unanswerable questions
  for {SQuAD}}. In \bibinfo{booktitle}{\emph{Association for Computational
  Linguistics}}.
\newblock


\bibitem[\protect\citeauthoryear{Rajpurkar, Zhang, Lopyrev, and
  Liang}{Rajpurkar et~al\mbox{.}}{2016}]%
        {squad2016}
\bibfield{author}{\bibinfo{person}{Pranav Rajpurkar}, \bibinfo{person}{Jian
  Zhang}, \bibinfo{person}{Konstantin Lopyrev}, {and} \bibinfo{person}{Percy
  Liang}.} \bibinfo{year}{2016}\natexlab{}.
\newblock \showarticletitle{SQuAD: 100,000+ questions for machine comprehension
  of text}. In \bibinfo{booktitle}{\emph{Empirical Methods in Natural Language
  Processing}}.
\newblock


\bibitem[\protect\citeauthoryear{Ratner, Bach, Ehrenberg, Fries, Wu, and
  R\'e}{Ratner et~al\mbox{.}}{2017}]%
        {snorkel-mainpaper}
\bibfield{author}{\bibinfo{person}{Alexander Ratner},
  \bibinfo{person}{Stephen~H. Bach}, \bibinfo{person}{Henry Ehrenberg},
  \bibinfo{person}{Jason Fries}, \bibinfo{person}{Sen Wu}, {and}
  \bibinfo{person}{Christopher R\'e}.} \bibinfo{year}{2017}\natexlab{}.
\newblock \showarticletitle{Snorkel: Rapid Training Data Creation with Weak
  Supervision}. In \bibinfo{booktitle}{\emph{Proceedings of the VLDB
  Endowment}} \emph{(\bibinfo{series}{VLDB'17})}.
\newblock


\bibitem[\protect\citeauthoryear{Rey and Neuh{\"a}user}{Rey and
  Neuh{\"a}user}{2011}]%
        {Rey2011}
\bibfield{author}{\bibinfo{person}{Denise Rey} {and} \bibinfo{person}{Markus
  Neuh{\"a}user}.} \bibinfo{year}{2011}\natexlab{}.
\newblock \bibinfo{booktitle}{\emph{Wilcoxon-Signed-Rank Test}}.
\newblock \bibinfo{publisher}{Springer Berlin Heidelberg},
  \bibinfo{address}{Berlin, Heidelberg}, \bibinfo{pages}{1658--1659}.
\newblock
\showISBNx{978-3-642-04898-2}
\urldef\tempurl%
\url{https://doi.org/10.1007/978-3-642-04898-2_616}
\showDOI{\tempurl}


\bibitem[\protect\citeauthoryear{Roth and Small}{Roth and Small}{2006}]%
        {al-application-structured-output-2006}
\bibfield{author}{\bibinfo{person}{Dan Roth} {and} \bibinfo{person}{Kevin
  Small}.} \bibinfo{year}{2006}\natexlab{}.
\newblock \showarticletitle{Margin-Based Active Learning for Structured Output
  Spaces}. In \bibinfo{booktitle}{\emph{Machine Learning: ECML 2006}},
  \bibfield{editor}{\bibinfo{person}{Johannes F{\"u}rnkranz},
  \bibinfo{person}{Tobias Scheffer}, {and} \bibinfo{person}{Myra Spiliopoulou}}
  (Eds.). \bibinfo{publisher}{Springer Berlin Heidelberg},
  \bibinfo{address}{Berlin, Heidelberg}, \bibinfo{pages}{413--424}.
\newblock
\showISBNx{978-3-540-46056-5}


\bibitem[\protect\citeauthoryear{Sanh, Debut, Chaumond, and Wolf}{Sanh
  et~al\mbox{.}}{2020}]%
        {sanh2020distilbert}
\bibfield{author}{\bibinfo{person}{Victor Sanh}, \bibinfo{person}{Lysandre
  Debut}, \bibinfo{person}{Julien Chaumond}, {and} \bibinfo{person}{Thomas
  Wolf}.} \bibinfo{year}{2020}\natexlab{}.
\newblock \bibinfo{title}{DistilBERT, a distilled version of BERT: smaller,
  faster, cheaper and lighter}.
\newblock
\newblock
\showeprint[arxiv]{1910.01108}~[cs.CL]


\bibitem[\protect\citeauthoryear{Scheffer, Decomain, and Wrobel}{Scheffer
  et~al\mbox{.}}{2001}]%
        {Scheffer01activehidden}
\bibfield{author}{\bibinfo{person}{Tobias Scheffer}, \bibinfo{person}{Christian
  Decomain}, {and} \bibinfo{person}{Stefan Wrobel}.}
  \bibinfo{year}{2001}\natexlab{}.
\newblock \bibinfo{title}{Active Hidden Markov Models for Information
  Extraction}.
\newblock
\newblock


\bibitem[\protect\citeauthoryear{Sener and Savarese}{Sener and
  Savarese}{2018}]%
        {sener2018active}
\bibfield{author}{\bibinfo{person}{Ozan Sener} {and} \bibinfo{person}{Silvio
  Savarese}.} \bibinfo{year}{2018}\natexlab{}.
\newblock \bibinfo{title}{Active Learning for Convolutional Neural Networks: A
  Core-Set Approach}.
\newblock
\newblock
\showeprint[arxiv]{1708.00489}~[stat.ML]


\bibitem[\protect\citeauthoryear{Settles}{Settles}{2009}]%
        {settles2009active}
\bibfield{author}{\bibinfo{person}{Burr Settles}.}
  \bibinfo{year}{2009}\natexlab{}.
\newblock \bibinfo{booktitle}{\emph{Active Learning Literature Survey}}.
\newblock \bibinfo{type}{Computer Sciences Technical Report} 1648.
  \bibinfo{institution}{University of Wisconsin--Madison}.
\newblock


\bibitem[\protect\citeauthoryear{Settles and Craven}{Settles and
  Craven}{2008}]%
        {variation-ratio-settles-2008}
\bibfield{author}{\bibinfo{person}{Burr Settles} {and} \bibinfo{person}{Mark
  Craven}.} \bibinfo{year}{2008}\natexlab{}.
\newblock \showarticletitle{An Analysis of Active Learning Strategies for
  Sequence Labeling Tasks} \emph{(\bibinfo{series}{EMNLP '08})}.
  \bibinfo{publisher}{Association for Computational Linguistics},
  \bibinfo{address}{USA}, \bibinfo{pages}{1070–1079}.
\newblock


\bibitem[\protect\citeauthoryear{Settles, Craven, and Ray}{Settles
  et~al\mbox{.}}{2007}]%
        {egl-settles-2008}
\bibfield{author}{\bibinfo{person}{Burr Settles}, \bibinfo{person}{Mark
  Craven}, {and} \bibinfo{person}{Soumya Ray}.}
  \bibinfo{year}{2007}\natexlab{}.
\newblock \showarticletitle{Multiple-Instance Active Learning}. In
  \bibinfo{booktitle}{\emph{Proceedings of the 20th International Conference on
  Neural Information Processing Systems}} (Vancouver, British Columbia, Canada)
  \emph{(\bibinfo{series}{NIPS'07})}. \bibinfo{publisher}{Curran Associates
  Inc.}, \bibinfo{address}{Red Hook, NY, USA}, \bibinfo{pages}{1289–1296}.
\newblock
\showISBNx{9781605603520}


\bibitem[\protect\citeauthoryear{Settles, Craven, and Ray}{Settles
  et~al\mbox{.}}{2008}]%
        {al-hybrid-2014}
\bibfield{author}{\bibinfo{person}{Burr Settles}, \bibinfo{person}{Mark
  Craven}, {and} \bibinfo{person}{Soumya Ray}.}
  \bibinfo{year}{2008}\natexlab{}.
\newblock \showarticletitle{Multiple-Instance Active Learning}. In
  \bibinfo{booktitle}{\emph{Advances in Neural Information Processing
  Systems}}, \bibfield{editor}{\bibinfo{person}{J.~Platt},
  \bibinfo{person}{D.~Koller}, \bibinfo{person}{Y.~Singer}, {and}
  \bibinfo{person}{S.~Roweis}} (Eds.), Vol.~\bibinfo{volume}{20}.
  \bibinfo{publisher}{Curran Associates, Inc.}
\newblock
\urldef\tempurl%
\url{https://proceedings.neurips.cc/paper/2007/file/a1519de5b5d44b31a01de013b9b51a80-Paper.pdf}
\showURL{%
\tempurl}


\bibitem[\protect\citeauthoryear{Siddhant and Lipton}{Siddhant and
  Lipton}{2018}]%
        {siddhant2018deep}
\bibfield{author}{\bibinfo{person}{Aditya Siddhant} {and}
  \bibinfo{person}{Zachary~C. Lipton}.} \bibinfo{year}{2018}\natexlab{}.
\newblock \showarticletitle{Deep {B}ayesian Active Learning for Natural
  Language Processing: Results of a Large-Scale Empirical Study}. In
  \bibinfo{booktitle}{\emph{Proceedings of the 2018 Conference on Empirical
  Methods in Natural Language Processing}}. \bibinfo{publisher}{Association for
  Computational Linguistics}, \bibinfo{address}{Brussels, Belgium},
  \bibinfo{pages}{2904--2909}.
\newblock
\urldef\tempurl%
\url{https://doi.org/10.18653/v1/D18-1318}
\showDOI{\tempurl}


\bibitem[\protect\citeauthoryear{Tomanek and Hahn}{Tomanek and Hahn}{2009}]%
        {al-application-structured-output-2009}
\bibfield{author}{\bibinfo{person}{Katrin Tomanek} {and} \bibinfo{person}{Udo
  Hahn}.} \bibinfo{year}{2009}\natexlab{}.
\newblock \showarticletitle{Reducing Class Imbalance during Active Learning for
  Named Entity Annotation}. In \bibinfo{booktitle}{\emph{Proceedings of the
  Fifth International Conference on Knowledge Capture}} (Redondo Beach,
  California, USA) \emph{(\bibinfo{series}{K-CAP '09})}.
  \bibinfo{publisher}{Association for Computing Machinery},
  \bibinfo{address}{New York, NY, USA}, \bibinfo{pages}{105–112}.
\newblock
\showISBNx{9781605586588}
\urldef\tempurl%
\url{https://doi.org/10.1145/1597735.1597754}
\showDOI{\tempurl}


\bibitem[\protect\citeauthoryear{Tong and Koller}{Tong and Koller}{2002}]%
        {al-application-text-classification-2009}
\bibfield{author}{\bibinfo{person}{Simon Tong} {and} \bibinfo{person}{Daphne
  Koller}.} \bibinfo{year}{2002}\natexlab{}.
\newblock \showarticletitle{Support Vector Machine Active Learning with
  Applications to Text Classification}.
\newblock \bibinfo{journal}{\emph{J. Mach. Learn. Res.}}  \bibinfo{volume}{2}
  (\bibinfo{date}{March} \bibinfo{year}{2002}), \bibinfo{pages}{45–66}.
\newblock
\showISSN{1532-4435}
\urldef\tempurl%
\url{https://doi.org/10.1162/153244302760185243}
\showDOI{\tempurl}


\bibitem[\protect\citeauthoryear{Wang, Singh, Michael, Hill, Levy, and
  Bowman}{Wang et~al\mbox{.}}{2018}]%
        {glue2018}
\bibfield{author}{\bibinfo{person}{Alex Wang}, \bibinfo{person}{Amanpreet
  Singh}, \bibinfo{person}{Julian Michael}, \bibinfo{person}{Felix Hill},
  \bibinfo{person}{Omer Levy}, {and} \bibinfo{person}{Samuel Bowman}.}
  \bibinfo{year}{2018}\natexlab{}.
\newblock \showarticletitle{GLUE: A Multi-Task Benchmark and Analysis Platform
  for Natural Language Understanding}. In \bibinfo{booktitle}{\emph{Empirical
  Methods in Natural Language Processing}}.
\newblock


\bibitem[\protect\citeauthoryear{Wei, Iyer, and Bilmes}{Wei
  et~al\mbox{.}}{2015}]%
        {submodularity-2015}
\bibfield{author}{\bibinfo{person}{Kai Wei}, \bibinfo{person}{Rishabh Iyer},
  {and} \bibinfo{person}{Jeff Bilmes}.} \bibinfo{year}{2015}\natexlab{}.
\newblock \showarticletitle{Submodularity in Data Subset Selection and Active
  Learning}. In \bibinfo{booktitle}{\emph{Proceedings of the 32nd International
  Conference on International Conference on Machine Learning - Volume 37}}
  (Lille, France) \emph{(\bibinfo{series}{ICML'15})}.
  \bibinfo{publisher}{JMLR.org}, \bibinfo{pages}{1954–1963}.
\newblock


\bibitem[\protect\citeauthoryear{Xu, Akella, and Zhang}{Xu
  et~al\mbox{.}}{2016}]%
        {Xu_incorporatingdiversity}
\bibfield{author}{\bibinfo{person}{Zuobing Xu}, \bibinfo{person}{Ram Akella},
  {and} \bibinfo{person}{Yi Zhang}.} \bibinfo{year}{2016}\natexlab{}.
\newblock \bibinfo{title}{Incorporating Diversity and Density in Active
  Learning for Relevance Feedback}.
\newblock
\newblock


\bibitem[\protect\citeauthoryear{Yinhan~Liu, Goyal, Du, Joshi, Chen, Levy,
  Lewis, Zettlemoyer, and Stoyanov}{Yinhan~Liu et~al\mbox{.}}{2019}]%
        {roberta2019}
\bibfield{author}{\bibinfo{person}{Myle~Ott Yinhan~Liu}, \bibinfo{person}{Naman
  Goyal}, \bibinfo{person}{Jingfei Du}, \bibinfo{person}{Mandar Joshi},
  \bibinfo{person}{Danqi Chen}, \bibinfo{person}{Omer Levy},
  \bibinfo{person}{Mike Lewis}, \bibinfo{person}{Luke Zettlemoyer}, {and}
  \bibinfo{person}{Veselin Stoyanov}.} \bibinfo{year}{2019}\natexlab{}.
\newblock \showarticletitle{RoBERTa: A Robustly Optimized BERT Pretraining
  Approach}.
\newblock \bibinfo{journal}{\emph{Computing Research Repository}}
  \bibinfo{volume}{arXiv:1907.11692} (\bibinfo{year}{2019}).
\newblock
\urldef\tempurl%
\url{https://arxiv.org/pdf/1907.11692.pdf}
\showURL{%
\tempurl}
\newblock
\shownote{version 1.}


\bibitem[\protect\citeauthoryear{Yoo and Kweon}{Yoo and Kweon}{2019}]%
        {yoo2019learning}
\bibfield{author}{\bibinfo{person}{Donggeun Yoo} {and} \bibinfo{person}{In~So
  Kweon}.} \bibinfo{year}{2019}\natexlab{}.
\newblock \showarticletitle{Learning Loss for Active Learning}. In
  \bibinfo{booktitle}{\emph{2019 IEEE/CVF Conference on Computer Vision and
  Pattern Recognition (CVPR)}}. \bibinfo{pages}{93--102}.
\newblock
\urldef\tempurl%
\url{https://doi.org/10.1109/CVPR.2019.00018}
\showDOI{\tempurl}


\bibitem[\protect\citeauthoryear{Zhang}{Zhang}{2019}]%
        {Zhang2019AnED}
\bibfield{author}{\bibinfo{person}{L. Zhang}.} \bibinfo{year}{2019}\natexlab{}.
\newblock \showarticletitle{An Ensemble Deep Active Learning Method for Intent
  Classification}.
\newblock \bibinfo{journal}{\emph{Proceedings of the 2019 3rd International
  Conference on Computer Science and Artificial Intelligence}}
  (\bibinfo{year}{2019}).
\newblock


\bibitem[\protect\citeauthoryear{Zhang, Zhao, and LeCun}{Zhang
  et~al\mbox{.}}{2015}]%
        {data-ag-2016}
\bibfield{author}{\bibinfo{person}{Xiang Zhang}, \bibinfo{person}{Junbo Zhao},
  {and} \bibinfo{person}{Yann LeCun}.} \bibinfo{year}{2015}\natexlab{}.
\newblock \showarticletitle{Character-Level Convolutional Networks for Text
  Classification}. In \bibinfo{booktitle}{\emph{Proceedings of the 28th
  International Conference on Neural Information Processing Systems - Volume
  1}} (Montreal, Canada) \emph{(\bibinfo{series}{NIPS'15})}.
  \bibinfo{publisher}{MIT Press}, \bibinfo{address}{Cambridge, MA, USA},
  \bibinfo{pages}{649–657}.
\newblock


\bibitem[\protect\citeauthoryear{Zhang, Lease, and Wallace}{Zhang
  et~al\mbox{.}}{2017}]%
        {zhang2016active}
\bibfield{author}{\bibinfo{person}{Ye Zhang}, \bibinfo{person}{Matthew Lease},
  {and} \bibinfo{person}{Byron~C. Wallace}.} \bibinfo{year}{2017}\natexlab{}.
\newblock \showarticletitle{Active Discriminative Text Representation
  Learning}. In \bibinfo{booktitle}{\emph{Proceedings of the Thirty-First AAAI
  Conference on Artificial Intelligence}} (San Francisco, California, USA)
  \emph{(\bibinfo{series}{AAAI'17})}. \bibinfo{publisher}{AAAI Press},
  \bibinfo{pages}{3386–3392}.
\newblock


\bibitem[\protect\citeauthoryear{Zhdanov}{Zhdanov}{2019}]%
        {zhdanov2019diverse}
\bibfield{author}{\bibinfo{person}{Fedor Zhdanov}.}
  \bibinfo{year}{2019}\natexlab{}.
\newblock \bibinfo{title}{Diverse mini-batch Active Learning}.
\newblock
\newblock
\showeprint[arxiv]{1901.05954}~[cs.LG]
\urldef\tempurl%
\url{https://arxiv.org/pdf/1901.05954.pdf}
\showURL{%
\tempurl}


\bibitem[\protect\citeauthoryear{Zhu, Wang, Yao, and Tsou}{Zhu
  et~al\mbox{.}}{2008}]%
        {entropy-uncertainty-2008}
\bibfield{author}{\bibinfo{person}{Jingbo Zhu}, \bibinfo{person}{Huizhen Wang},
  \bibinfo{person}{Tianshun Yao}, {and} \bibinfo{person}{Benjamin~K. Tsou}.}
  \bibinfo{year}{2008}\natexlab{}.
\newblock \showarticletitle{Active Learning with Sampling by Uncertainty and
  Density for Word Sense Disambiguation and Text Classification}. In
  \bibinfo{booktitle}{\emph{Proceedings of the 22nd International Conference on
  Computational Linguistics - Volume 1}} (Manchester, United Kingdom)
  \emph{(\bibinfo{series}{COLING'08})}. \bibinfo{publisher}{Association for
  Computational Linguistics}, \bibinfo{address}{USA},
  \bibinfo{pages}{1137–1144}.
\newblock
\showISBNx{9781905593446}


\end{thebibliography}

\end{document}